\documentclass{jpp}
\usepackage{graphicx}

\usepackage[utf8]{inputenc}
\usepackage[T1]{fontenc}
\usepackage{amsmath}

\usepackage{xcolor}
\usepackage{soul}

\shorttitle{Available energy}
\shortauthor{Helander and Mackenbach}

\title{Available energy of plasmas with small fluctuations}

\author{P.~Helander\aff{1}
  \corresp{\email{per.helander@ipp.mpg.de}} 
 \and R.J.J.~Mackenbach\aff{2,3}}

\affiliation{\aff{1}Max Planck Institute for Plasma Physics, Greifswald, Germany
\aff{2}École Polytechnique Fédérale de Lausanne (EPFL), Swiss Plasma Center (SPC), CH-1015 Lausanne, Switzerland
\aff{3}Eindhoven University of Technology, Eindhoven, Netherlands}

\begin{document}

\maketitle

\begin{abstract}
The available energy of a plasma is defined as the maximum amount by which the plasma energy can be lowered by volume-preserving rearrangements in phase space, a so-called Gardner re-stacking. A general expression is derived for the available energy of a nearly homogeneous plasma and is shown to be closely related to the Helmholtz free energy, which it can never exceed. A number of explicit examples are given. 

\end{abstract}

\section{Introduction}

Two previous publications \citep{helander_2017,helander_2020} in this journal have discussed the `available energy' of a collisionless plasma, defined as the largest amount by which the total kinetic energy of the plasma particles can be lowered by any motion subject to the constraints that follow from the conservation of phase-space volume and adiabatic invariants. It was proposed that this quantity could serve as a measure of nonlinear stability, and it was found that it can indeed be useful for deriving stability criteria. Moreover, the available energy of trapped electrons has more recently been shown to be correlated with the energy flux in density-gradient-driven turbulence computed by gyrokinetic simulations \citep{mackenbach_2022,Mackenbach_Proll_Wakelkamp_Helander_2023}. Different types of stellarators and tokamaks possess quite different amounts of available energy with respect to instabilities and turbulence that preserve the magnetic moment and parallel adiabatic invariant of electrons, and these differences are reflected in the turbulent energy flux. In particular, the degree to which regions with low magnetic field strength containing magnetically trapped particles overlap with regions of unfavourable magnetic curvature (convex field lines) varies greatly between different magnetic-confinement devices, leading to large differences in available energy and trapped-electron-mode turbulence. In tokamaks, the regions of magnetic trapping and unfavourable (or `bad') curvature overlap almost perfectly, both being situated on the outboard side of the torus, making the available energy relatively large. In contrast, in the Wendelstein 7-X stellarator \citep{wolf_2019}, the worst curvature is found in regions with relatively high field strength, where there are particularly few trapped particles, making the available energy small. As a result, trapped-electron modes are much more stable in Wendelstein 7-X than in tokamaks \citep{proll_2013,helander_2015}. 

The correlation between available energy and turbulent transport has to do with the fact that the available energy measures the maximum amount of energy that can be converted into turbulent motion in plasma without energy input from the surroundings. Consider, for example, a plasma occupying a finite region of space with an insulating boundary. If the distribution function at some initial instant is a Maxwellian with constant density, constant temperature, and zero mean velocity, the available energy vanishes and the plasma is clearly stable. If the initial density, temperature or flow velocity instead vary slightly, there is some available energy present in the system, which can be converted into kinetic energy in turbulent eddies and cause a flux of particles and energy across the domain (but of course not across the insulating boundary). 

The notion of available energy thus bears resemblance to the concept of free energy in thermodynamics, which quantifies how much of a system's internal energy can be converted to work and thus to kinetic energy. In the present paper, we clarify how available energy is related to free energy. We do so by considering the case of an almost homogeneous plasma and deriving a general formula for the available energy, which can be compared with that for the Helmholtz free energy. As we shall see, the latter is an absolute upper bound on the available energy. 

\section{Basic notation and equations}

We follow the notation of \cite{helander_2017,helander_2020} and denote the phase-space coordinates by the vector $\boldsymbol{ x}$ and their Jacobian by $\sqrt{g(\boldsymbol{ x})}$, so that the phase-space volume element is $\sqrt{g} d\boldsymbol{x}$. If adiabatic invariants $\boldsymbol{ y}$ are conserved, some of the coordinates $\boldsymbol{ x}$ will be chosen to be equal to these while the remaining ones are denoted by $\boldsymbol{ z}$, so that $\boldsymbol{ x} = (\boldsymbol{ y}, \boldsymbol{ z})$. The particle energy is some function of the phase-space coordinates, which we denote by $\epsilon(\boldsymbol{ x})$ and is typically equal to $\epsilon = mv^2/2$. The volume of the region of phase space at constant $\boldsymbol{y}$ in which the energy is at most equal to some value $w$ is 
    $$ \Omega(w,\boldsymbol{ y}) = \int \Theta[w - \epsilon(\boldsymbol{ y},\boldsymbol{ z})] \sqrt{g(\boldsymbol{ y},\boldsymbol{ z})} \; d\boldsymbol{ z}, $$
where $\Theta$ denotes the Heaviside step function. 

In this article, we only consider a single species of plasma particles, which are distributed in phase space according to some distribution function $f(\boldsymbol{ x},t)$. The total energy of all these particles is thus equal to
    $$ E(t) = \int \epsilon(\boldsymbol{ x}) f(\boldsymbol{ x},t) \sqrt{g(\boldsymbol{ x})} \; d\boldsymbol{ x}. $$
As the distribution function evolves in time in accordance with the Vlasov equation or any other collisionless kinetic equation describing particles whose motion is Hamiltonian, the flow in phase space is incompressible. As a result, there is a lower limit to the energy $E(t)$ \citep{Gardner-1963,kolmes_2020}, and its lowest possible value is given by 
    $$E_0 = \int \epsilon f_0 \sqrt{g} \; d\boldsymbol{ x}, $$
where $f_0$ is a distribution function that depends on the arguments as $f_0(\boldsymbol{ y},\boldsymbol{ z}) =F_0[\epsilon(\boldsymbol{ y},\boldsymbol{ z}),\boldsymbol{ y}]$.\footnote{Although the motion of all particles is assumed to be Hamiltonian, their total energy need not be conserved. Energy could, for instance, be transferred from the particles to the electric field.} As shown by \citet{helander_2017,helander_2020}, $F_0$ is a monotonically decreasing function of the first argument that is determined by the integral equation
    \begin{equation}
    \int \Theta[f(\boldsymbol{ y},\boldsymbol{ z}) - F_0(w,\boldsymbol{ y}) ] \sqrt{g(\boldsymbol{ y},\boldsymbol{ z})} \; d\boldsymbol{ z} = \Omega(w,\boldsymbol{ y}), 
    \label{ground state eq}
    \end{equation}
where the variables $\boldsymbol{y}$ are held constant in the integration. The function $f_0$ represents a minimum-energy state of the species in question, the so-called ground state. The available energy is the difference between the initial energy and the ground-state energy,
    $$ A = \int (f-f_0) \epsilon \sqrt{g} \; d\boldsymbol{ x}.$$
It is important to note that although the ground state is uniquely defined by the initial state, different initial states in general correspond to different ground states. 

\section{Available energy close to a ground state}

\subsection{Ground state corresponding to a given perturbation}

We now turn to the central question in this paper and calculate the energy available to a plasma close to a ground state. To this end, suppose that 
    $$ f_0(\boldsymbol{ x}) = F_0[\epsilon(\boldsymbol{ x}),\boldsymbol{ y}] $$
denotes a ground state and
    $$ f(\boldsymbol{ x}) = f_0(\boldsymbol{ x}) + \delta f(\boldsymbol{ x}) $$
a nearby state, where $\delta f \ll f_0$. Note that, although $f_0$ is a ground state, it is in general inaccessible to a plasma with the distribution function $f$. We thus need to calculate the ground state corresponding to $f$, which we write as
    $$ F[\epsilon(\boldsymbol{ x}),\boldsymbol{ y}] = 
    F_0[\epsilon(\boldsymbol{ x}),\boldsymbol{ y}] + \delta F[\epsilon(\boldsymbol{ x}),\boldsymbol{ y}], $$
where we expect $\delta F \ll F_0$. (As we shall discuss later, the derivatives of $\delta F$ also need to be much smaller than those of $F_0$.) In the interest of economy, we write $\boldsymbol{ x}$ instead of $(\boldsymbol{ y},\boldsymbol{ z})$ wherever possible. The functions $F_0$ and $\delta F$ are then defined by the integral equations    
    $$ \int \Theta[f_0(\boldsymbol{ x}) - F_0(w,\boldsymbol{ y}) ] \sqrt{g} \; d\boldsymbol{ z} = \Omega(w,\boldsymbol{ y}),$$
    $$ \int \Theta[f_0(\boldsymbol{ x}) + \delta f(\boldsymbol{ x}) - F_0(w,\boldsymbol{ y}) - \delta F (w,\boldsymbol{ y})] 
    \sqrt{g} \; d\boldsymbol{ z} = \Omega(w,\boldsymbol{ y}). $$
The last equation is now expanded to second order, giving
    $$ \int \left\{ \Theta[f_0(\boldsymbol{ x}) - F_0(w,\boldsymbol{ y}) ] + [\delta f(\boldsymbol{ x}) - \delta F (w,\boldsymbol{ y})]
    \delta [f_0(\boldsymbol{ x}) - F_0(w,\boldsymbol{ y}) ] \right.
     $$
    \begin{equation}
     \left.  + \frac{1}{2} [\delta f(\boldsymbol{ x})  - \delta F (w,\boldsymbol{ y})]^2 
    \delta' [f_0(\boldsymbol{ x}) - F_0(w,\boldsymbol{ y}) ] \right\} \sqrt{g}  \; d\boldsymbol{ z} = \Omega(w,\boldsymbol{ y}). 
    \label{2nd order}
        \end{equation}
Since $F_0(w,\boldsymbol{ y})$ is a decreasing function of $w$, we have the relation 
    $$ \Theta[w-\epsilon(\boldsymbol{ x}) ] = \Theta[f_0(\boldsymbol{ x}) - F_0(w,\boldsymbol{ y})], $$
which can be differentiated with respect to $w$ to give
    $$ \delta[f_0(\boldsymbol{ x}) - F_0(w,\boldsymbol{ y})] = - \frac{\delta[w-\epsilon(\boldsymbol{ x}) ]}{F_0'(w,\boldsymbol{ y})}, $$
where a prime denotes the derivative with respect to the first argument. Differentiating once more gives
    $$ \delta'[f_0(\boldsymbol{ x}) - F_0(w,\boldsymbol{ y})] 
    = \frac{1}{F_0'(w,\boldsymbol{ y})} \frac{\partial}{\partial w} \left(
    \frac{\delta[w-\epsilon(\boldsymbol{ x}) ]}{F'_0(w,\boldsymbol{ y})} \right). $$
Substituting these last two equations in (\ref{2nd order}) gives
	$$ \int \left\{ \delta[w-\epsilon(\boldsymbol{ x}) ][\delta f(\boldsymbol{ x}) - \delta F (w,\boldsymbol{ y})]
	- \frac{1}{2} [\delta f(\boldsymbol{ x}) - \delta F (w,\boldsymbol{ y})]^2 
	\frac{\partial}{\partial w} \left(
    \frac{\delta[w-\epsilon(\boldsymbol{ x}) ]}{F'_0(w,\boldsymbol{ y})} \right) \right\} \sqrt{g}  \; d\boldsymbol{ z} = 0, $$
where we use
	$$ \Omega'(w,\boldsymbol{ y}) = \int \delta[w-\epsilon(\boldsymbol{ x}) ] \sqrt{g}  \; d\boldsymbol{ z} $$
to conclude that
	\begin{equation} \delta F (w,\boldsymbol{ y}) = \frac{1}{\Omega'(w,\boldsymbol{ y})} \int 
	\left\{ \delta f(\boldsymbol{ x}) \delta[w-\epsilon(\boldsymbol{ x}) ]
	- \frac{1}{2} [\delta f(\boldsymbol{ x}) - \delta F (w,\boldsymbol{ y})]^2 
	\frac{\partial}{\partial w} \left(
    \frac{\delta[w-\epsilon(\boldsymbol{ x}) ]}{F'_0(w,\boldsymbol{ y})} \right) \right\} \sqrt{g}  \; d\boldsymbol{ z}, 
	\label{delta F}
	\end{equation}
The quantity $\Omega'(w,\boldsymbol{ y})$ is equal to the density of states in classical statistical mechanics. 
	
\subsection{Available energy}
 
Having thus derived an expression for the ground state that is accurate to second order, we now turn our attention to the available energy
	$$ A = \int \epsilon (\boldsymbol{ x}) (\delta f(\boldsymbol{ x}) - \delta F [\epsilon(\boldsymbol{ x}),\boldsymbol{ y}])	\sqrt{g} \; d\boldsymbol{ x} $$
	\begin{equation} =  \int \epsilon (\boldsymbol{ x}) \delta f(\boldsymbol{ x}) \sqrt{g} \; d\boldsymbol{ x}
	- \int d\boldsymbol{ y} \int_0^\infty \delta F (w,\boldsymbol{ y}) w dw   \int \delta[w-\epsilon(\boldsymbol{ x}) ]
	\sqrt{g} \; d\boldsymbol{ z}, 
 \label{A-calculation}
 \end{equation}
which is ostensibly of first order in the smallness of $\delta f$. However, when (\ref{delta F}) is substituted for $\delta F$, the first-order terms cancel, since in leading order
\begin{equation}
\int \delta F(w,\boldsymbol{ y}) \delta[w - \epsilon(\boldsymbol{ x})]\sqrt{g} \: d \boldsymbol{ z} \simeq \int \delta f(\boldsymbol{ x}) \delta [ w - \epsilon(\boldsymbol{ x})] \sqrt{g} \: d \boldsymbol{ z}.
\label{eq: delta F leading order integral}
\end{equation}
When substituted in Eq.~(\ref{A-calculation}), this result cancels the first term on the right, and the available energy thus vanishes in this order. We therefore continue to the next order, where
	$$ A = \frac{1}{2} \int \sqrt{g} \; d\boldsymbol{ x} \int_0^\infty w [\delta f(\boldsymbol{ x}) - \delta F (w,\boldsymbol{ y})]^2
	\frac{\partial}{\partial w} \left(
    \frac{\delta[w-\epsilon(\boldsymbol{ x}) ]}{F'_0(w,\boldsymbol{ y})} \right) dw. $$
Focusing on the integral over $w$, one can make progress by integrating by parts,
    $$  \int_0^\infty w [\delta f(\boldsymbol{ x}) - \delta F (w,\boldsymbol{ y})]^2
	\frac{\partial}{\partial w} \left(
    \frac{\delta[w-\epsilon(\boldsymbol{ x}) ]}{F'_0(w,\boldsymbol{ y})} \right) dw $$
    $$= - \int_0^\infty \left\{ [\delta f(\boldsymbol{ x}) - \delta F (w,\boldsymbol{ y})]^2 - 2 w [\delta f(\boldsymbol{ x}) - \delta F (w,\boldsymbol{ y})] \delta F '(w,\boldsymbol{ y}) \right\} \frac{\delta[w-\epsilon(\boldsymbol{ x}) ]}{F'_0(w,\boldsymbol{ y})} \: dw, $$ 
and the available energy becomes
\begin{equation}
    \begin{aligned}
        A =& - \frac{1}{2} \int_0^\infty dw \int \frac{d\boldsymbol{ y}}{F'_0(w,\boldsymbol{ y})}
	        \\
           & \int \left\{[\delta f(\boldsymbol{ x}) - \delta F (w,\boldsymbol{ y})]^2 
	-2 w [\delta f(\boldsymbol{ x}) - \delta F (w,\boldsymbol{ y})] \delta F' (w,\boldsymbol{ y}) \right\} 
	\delta[w-\epsilon(\boldsymbol{ x}) ] \sqrt{g} \; d\boldsymbol{ z}.
    \end{aligned}
\end{equation}
Since this expression is of second order, the first-order version of (\ref{2nd order}) may now be used for $\delta F$, 
	\begin{equation} \delta F (w,\boldsymbol{ y}) = \frac{1}{\Omega'(w,\boldsymbol{ y})} \int 
	\delta f(\boldsymbol{ x}) \delta[w-\epsilon(\boldsymbol{ x}) ] \sqrt{g} \; d\boldsymbol{ z},
   \label{approximate delta F}
	\end{equation}
whereupon the last term vanishes as a consequence of Eq. (\ref{eq: delta F leading order integral}),
	$$ \int_0^\infty w dw \int \frac{\delta F'(w,\boldsymbol{ y})}{F'_0(w,\boldsymbol{ y})} d\boldsymbol{ y} 
	\int [\delta f(\boldsymbol{ x}) - \delta F (w ,\boldsymbol{ y})] \delta[w-\epsilon(\boldsymbol{ x}) ] \sqrt{g} \; d\boldsymbol{ z} = 0. $$
We thus obtain the following expression for the available energy,
\begin{equation} 
	A = - \int \frac{(\delta f(\boldsymbol{ x}) - \delta F [\epsilon(\boldsymbol{ x}),\boldsymbol{ y}])^2}{2 F'_0[\epsilon(\boldsymbol{ x}),\boldsymbol{ y}]} 
	\sqrt{g} \; d\boldsymbol{ x}, 
	\label{A1}
\end{equation}
where $\delta F$ is given by (\ref{approximate delta F}). Since $F'_0(w,\boldsymbol{ y})$ is negative, $A$ is positive  definite. Moreover, it vanishes if and only if $\delta f(\boldsymbol{ y},\boldsymbol{ z}) = \delta F [\epsilon(\boldsymbol{ y},\boldsymbol{ z}),\boldsymbol{ y}]$, so that $f = f_0 + \delta f$ only depends on $\boldsymbol{ z}$ through the energy function $\epsilon$, which is, of course, the condition that $f$ represents a ground state. 

Another useful form for the available energy is obtained by expanding the square in (\ref{A1}) and noting that, to leading order, 
	$$ \int \frac{\delta f(\boldsymbol{ x}) \delta F [\epsilon(\boldsymbol{ x}),\boldsymbol{ y}]}{2 F'_0[\epsilon(\boldsymbol{x}),\boldsymbol{ y}]} 
	\sqrt{g} \; d\boldsymbol{ x}
	\simeq \int \frac{\delta F^2 [\epsilon(\boldsymbol{ x}),\boldsymbol{ y}]}{2 F'_0[\epsilon(\boldsymbol{x}),\boldsymbol{ y}]} 
	\sqrt{g} \; d\boldsymbol{ x}, $$
as follows from  
	$$ \int \frac{\delta f(\boldsymbol{ x}) \delta F [\epsilon(\boldsymbol{ x}),\boldsymbol{ y}]}{2 F'_0[\epsilon(\boldsymbol{x}),\boldsymbol{ y}]} 
	\sqrt{g} \; d\boldsymbol{ x}
	= \int_0^\infty dw \int \frac{\delta f(\boldsymbol{ x}) \delta F(w,\boldsymbol{ y})}{2 F'_0(w,\boldsymbol{ y})} 
	\delta[w-\epsilon(\boldsymbol{ x}) ]\sqrt{g} \; d\boldsymbol{ y} d\boldsymbol{ z},
	$$
where substituting Eq. (\ref{eq: delta F leading order integral}) gives the desired result. We thus find, to the same accuracy as Eq.~(\ref{A1}),
	\begin{equation} 
	A = - \int \frac{\delta f^2(\boldsymbol{ x}) - \delta F^2 [\epsilon(\boldsymbol{ x}),\boldsymbol{ y}]}{2 F'_0[\epsilon(\boldsymbol{ x}),\boldsymbol{ y}]} 
	\sqrt{g} \; d\boldsymbol{ x}.
	\label{A2}
	\end{equation}

\subsection{Relation to Helmholtz free energy}

Let us temporarily assume that $f_0$ is a Maxwellian with spatially constant density $n_0$ and temperature $T_0$, 
	$$ f_0 = n_0 \left( \frac{m}{2 \pi T_0} \right)^{3/2} e^{-\epsilon / T_0}, $$
so that $F'_0 = - F_0/T_0$, and let us take the total number of particles contained in the distribution functions $f_0$ and $f$ to be the same,
	$$ \int f_0 \sqrt{g} \; d\boldsymbol{ x} = \int f \sqrt{g} \; d\boldsymbol{ x}. $$
If we define the difference in entropy and energy carried by the distribution functions $f_0$ and $f$ as
	$$ \delta S = - \int (f \ln f - f_0 \ln f_0) \sqrt{g} \; d\boldsymbol{ x}, $$
	$$ \delta U = \int \epsilon(\boldsymbol{ x}) (f-f_0) \sqrt{g} \; d\boldsymbol{ x}, $$
then to second order in $\delta f$
	$$ \delta S = \frac{\delta U}{T_0} - \int \frac{\delta f^2}{2 f_0} \sqrt{g} \; d\boldsymbol{ x}, $$
and we conclude that 
	\begin{equation}
	H = T_0 \int \frac{\delta f^2}{2 f_0} \sqrt{g} \; d\boldsymbol{ x}  = \delta U - T_0 \delta S
	\label{H}
	\end{equation}
denotes the difference in Helmholtz free energy in the two distribution functions. In plasma physics, this quantity has often been considered in discussions of turbulence, see e.g. \cite{Krommes-1993,Brizard-1994,Sugama-1996,Garbet-2005,Schekochihin-2009,Banon-2011,Stoltzfus-Dueck-2017}. It has recently been used to derive upper bounds on gyrokinetic instabilities that are generally valid irrespective of the geometry of the magnetic field, the number of particle species, collisions etc. \citep{helander_2021,helander_plunk_2022,plunk_helander_2022}

Equations (\ref{A1}) and (\ref{A2}) show that the available energy is closely related to, but in general different from, the Helmholtz free energy. If for some reason $\delta F$ vanishes, then these two energies are equal, i.e. $H=A$ if $\delta F = 0$, and otherwise the available energy is smaller than the Helmholtz free energy, i.e. $A < H$ if $\delta F \ne 0$. 

This line of thought can be extended to the case of a non-Maxwellian $f_0$ by considering a more general entropy functional
    $$ S[f] = \int s(f) \sqrt{g} \; d\boldsymbol{ x}, $$
where the function $s$ will be chosen suitably. To second order in $\delta f / f_0 $ we then have
    $$ \delta S = S[f] - S[f_0] = \int \left[ s'(f_0) \delta f  + \frac{s''(f_0)}{2}  \delta f^2\right]\sqrt{g} \; d\boldsymbol{ x}, $$
and the natural generalisation of the Helmholtz energy is
    $$ H = T_0 \int \left[ \left( \frac{\epsilon}{T_0} - s'(f_0) \right) \delta f 
    - \frac{s''(f_0)}{2}  \delta f^2\right] \sqrt{g} \; d\boldsymbol{ x}, $$
if the function $s$ and the constant $T_0$ are chosen such that the linear term vanishes. If $f_0({\bf x}) = F_0[\epsilon({\bf x})]$, this is accomplished by the choice
    $$ s'(F_0) = \frac{\epsilon}{T_0}, $$
    $$ s''(F_0) = \frac{1}{T_0 F_0'(\epsilon)}. $$
The resulting Helmholtz energy,     
    \begin{equation} H = - \int \frac{\delta f^2(\boldsymbol{ x})}{2 F'_0[\epsilon(\boldsymbol{ x})]} 
	\sqrt{g} \; d\boldsymbol{ x},
    \label{JBT H}
    \end{equation}
coincides with Eqs.~(\ref{A1}) and (\ref{A2}) if $\delta F = 0$. A similar argument was made by \cite{Taylor-1963}, see also \cite{Kruskal-oberman}.

This construction shows that, if $F_0$ is chosen in such a way that $\delta F = (\ref{approximate delta F})$ vanishes, then an appropriate definition of the entropy function $s(f)$ results in a Helmholtz energy that is equal to the available energy. In this sense, the available energy is always equal to the Helmholtz free energy with a suitable choice for the unperturbed state and the entropy. However, in practice, a basic requirement of perturbation theory is that the unperturbed state, in this case $F_0$, should be simple, which will usually not be the case if one insists that Eq.~(\ref{approximate delta F}) vanishes. 

The reason why it is necessary to include $\delta F$ is that, given a ground state $f_0$ and a nearby state $f = f_0 + \delta f$, the former is generally not accessible from the latter through Gardner restacking. The ground state corresponding to $f$ is instead $F = F_0 + \delta F$, and the available energy (\ref{A1}) is equal to the generalised free energy associated with a perturbation from {\em this} ground state rather than $f_0$. This circumstance also explains why the available energy cannot exceed the generalised Helmholtz free energy (\ref{JBT H}). The latter energy would be released through Gardner restacking of $f$ into $f_0$, if this were possible. Imposing the extra constraint that the ground state reached by restacking should be accessible from the initial state can only make the released energy smaller.  

\section{Available energy in special cases}

In this section we illustrate the practical utility of Eqs.~(\ref{A1}) and (\ref{A2}) by calculating the available energy in several special cases. 

\subsection{Bi-Maxwellian with fluctuating density, temperatures, and flow} \label{sec: Bi-Maxwellian with fluctuating density, temperatures, and flow}
We begin by considering the case of a bi-Maxwellian distribution function, i.e., 
    \begin{equation}
     f(\boldsymbol{ r}, \boldsymbol{ v}) = n \left( \frac{m}{2 \pi \overline{T}} \right)^{3/2} 
	\exp \left(-\frac{m (v_x-u_x)^2}{2T_\perp} -\frac{m (v_y- u_y)^2}{2T_\perp} - \frac{m (v_\parallel-u_\|)^2}{2T_\parallel}\right),
 \label{eq: bimaxwellian}
 \end{equation}
in Cartesian velocity-space coordinates aligned with the temperature anisotropy. The quantity $\overline{T} = T_\|^{1/3}T_\perp^{2/3}$ is the geometric mean of the temperatures in the three directions, two of which have been taken to be equal. The density, temperatures, and flow velocity, are assumed to be nearly constant
    $$n = n_0[1 + \nu(\boldsymbol{ r})], $$
    $$T_\| = T_{\|,0}[1 + \tau_\|(\boldsymbol{ r})],$$
    $$T_\perp = T_{\perp,0}[1 + \tau_\perp(\boldsymbol{ r})],$$
    $$  \boldsymbol{u} = \boldsymbol{u}_0 + \delta \boldsymbol{u}(\boldsymbol{r}),$$
where $ \boldsymbol{u}_0 $ can be made to vanish by a Galilean transformation. 
\cite{helander_2017} treated the case of constant density and temperatures and showed that the ground state is an isotropic Maxwellian with temperature $T_{\perp,0}=T_{\|,0}=T_0$.  The perturbed distribution function thus becomes $f = f_0 + \delta f$, with
\begin{equation}
    \delta f = \left[\nu + \tau_\perp \left( \frac{m v_\perp^2}{2 T_0} - 1 \right) + \tau_{\|} \left( \frac{m v_\|^2}{2 T_0} - \frac{1}{2} \right) - \frac{m (\boldsymbol{v} \cdot \delta \boldsymbol{u})}{T_0} \right] f_0(\boldsymbol{ r}, \boldsymbol{ v}),
    \label{eq: delta f bimaxwellian}
\end{equation}
where $\boldsymbol{v} = (v_x,v_y,v_\|)$ and $v_\perp^2 = v_x^2 + v_y^2$. The function $f_0$ can be chosen so that the distribution functions $f_0$ and $f$ carry equal amounts of particles and energy, 
    $$ \langle \nu \rangle = \left\langle \frac{\overline{T}-T_0}{T_0} \right\rangle = 0, $$
where angular brackets denote an average over the plasma volume $V$, 
\begin{equation}
    \langle \cdots \rangle = \frac{1}{V} \int \cdots \; d \boldsymbol{ r},
\end{equation}
To leading order we must then have $\langle \tau_\| \rangle = -2 \langle \tau_\perp \rangle$, and Eq.~(\ref{approximate delta F}) becomes
	$$ \delta F(w) = \frac{1}{\Omega'(w)} \int f_0 \tau_\perp
	\left(\epsilon - \frac{3}{2}\frac{m v_\|^2}{ T_0}  \right)
	\delta (w - \epsilon) d\boldsymbol{ r} d\boldsymbol{ v} . $$
One can make progress by introducing the coordinates $v_\perp = v \cos \vartheta$ and $v_\| = v \sin \vartheta$. The volume element becomes $d \boldsymbol{ v} = \pi (2/m)^{3/2} \cos \vartheta \sqrt{\epsilon} \; d \vartheta d \epsilon$, resulting in
    $$    \delta F(w) = \frac{\pi}{\Omega'(w)} \int f_0 \tau_\perp \left(\frac{2\epsilon}{m} \right)^{3/2}  \delta (w - \epsilon) \left( 1 - 3 \sin^2 \vartheta  \right) \cos \vartheta \; d \vartheta  d \epsilon d\boldsymbol{ r} = 0. $$
The available energy is thus equal to the Helmholtz free energy and becomes
	\begin{equation} A = H = \frac{n_0 T_0 V}{2} 
	\left\langle \nu^2 + \frac{1}{2}  \tau_\|^2 + \tau_\perp^2 + \frac{m (\delta \boldsymbol{u})^2}{T_0} \right\rangle.
    \label{eq: AE bimaxwellian no invariants}
 \end{equation}
This expression generalises the result found by \citet{helander_2017} to include anisotropic temperature perturbations and a spatially varying flow.

\subsection{Kappa distribution in any number of dimensions} \label{sec: kappa distribution in any dimension}
We next consider the kappa distribution in any integer number $d$ of dimensions. These functions were originally introduced to model high-energy tails of distribution functions in astrophysical plasmas that exhibit power-law behaviour \citep{olbert1968summary,vasyliunas1968survey}. \citet{lazar2021kappa} provide a recent review of their use in modelling various plasmas. The kappa distribution distribution function is defined by \citep{livadiotis2013understanding}
\begin{equation}
    f_{\kappa,d}(\boldsymbol{r},\boldsymbol{v}) = n \left(\frac{m}{2 \pi T}\right)^{d/2} \frac{\Gamma\left( \kappa_0 + 1 + \frac{d}{2} \right)}{\kappa_0^{d/2} \Gamma\left( \kappa_0 + 1 \right)} \left( 1 + \frac{1}{\kappa_0} \frac{mv^2}{2T} \right)^{-\kappa_0 - 1 - d/2},
\end{equation}
where $\kappa_0 = \kappa - d/2$, and the normalisation constants have been chosen such that the integral over velocity-space returns the number density, 
    $$\int f_{\kappa,d} d \boldsymbol{v} = n, $$
and the energy density is
    $$\int \frac{mv^2}{2} f_{\kappa,d} d \boldsymbol{v} = \frac{nTd}{2}, $$
so that each degree of freedom as usual carries the energy $nT/2$. Note that the tail of the distribution has a power-law behaviour, $f_{\kappa,d} \propto v^{-2\kappa_0 - 2 - d}$. Furthermore, the limit of $\kappa_0 \rightarrow \infty$ corresponds to the usual Maxwellian. \par 
Any spatially constant kappa distribution is trivially a ground state, as it is a monotonically decreasing function of $v^2$. If the density and temperature are perturbed by relative amounts $\nu(\boldsymbol{r})$ and $\tau(\boldsymbol{r})$, respectively, the fluctuating part of the distribution function becomes
\begin{equation}
    \delta f_{\kappa,d} = f_{\kappa,d,0} \left[ \nu + \tau \frac{\frac{m v^2}{2 T_0}(1 + \kappa_0) - \frac{\kappa_0 d}{2}}{\frac{m v^2}{2 T_0} + \kappa_0} \right].
\end{equation}
Imposing the condition $\langle \nu \rangle = \langle \tau \rangle = 0$, we find that the ground state vanishes, $\delta F = 0$. The available energy is thus equal to the Helmholtz free energy and is given by
\begin{equation}
    A = H = \frac{n_0 T_0 V}{2} \left \langle \frac{d( \nu + \tau )^2 + \kappa_0( 2 \nu^2 + d\tau^2 ) }{2 + d + 2 \kappa_0} \right \rangle.
\end{equation}
In the limit of a Maxwellian, $\kappa_0 \rightarrow \infty$, the available energy depends on the dimensionality $d$ as
$$
A = \frac{n_0 T_0 V}{2} \left \langle\nu^2 + \frac{d}{2} \tau^2 \right \rangle.
$$
Setting $d=3$ we recover the result of \cite{helander_2017} for the available energy of a three-dimensional Maxwellian plasma with slightly varying density and temperature.

\subsection{Magnetic-moment conservation of a bi-Maxwellian}
Let us again consider the case of a bi-Maxwellian at rest with slightly varying density and temperatures, where we now impose the condition that $\mu= mv_\perp^2/2B$ be conserved, so that Eq. (\ref{eq: bimaxwellian}) becomes
\begin{equation}
    f(\boldsymbol{ r}, \boldsymbol{ v}) = n(\boldsymbol{ r}) \left( \frac{m}{2 \pi \overline{T}(\boldsymbol{ r})} \right)^{3/2} 
	e^{- \epsilon / T_\parallel(\boldsymbol{ r}) + \mu B(\boldsymbol{ r}) [ 1/T_\parallel(\boldsymbol{ r}) -  1/T_\perp(\boldsymbol{ r})]},
 \label{eq: bimaxwellian with mu}
\end{equation}
where $\overline{T}$ is the geometric mean of the temperatures, as in Sec. \ref{sec: Bi-Maxwellian with fluctuating density, temperatures, and flow}. When the magnetic moment $\mu$ is conserved, any ground state must be of the form $f_0(\boldsymbol{ r}, \boldsymbol{v}) = F_0[\epsilon(\boldsymbol{ r}, \boldsymbol{ v}), \mu(\boldsymbol{ r}, \boldsymbol{ v})]$ with $F'_0(w,\mu) \le 0$. (As usual, a prime denotes differentiation with respect to the first argument.) It follows that, in order for the function 
$$f_0=n_0 \left( \frac{m}{2 \pi \overline{T}_0} \right)^{3/2} 
	e^{- \epsilon / T_{\parallel,0} + \mu B(\boldsymbol{ r}) [ 1/T_{\parallel,0} -  1/T_{\perp,0}]}$$ 
to be a ground state, the perpendicular and parallel temperatures either must be equal $T_{\perp,0}=T_{\|,0} = T_0$, if $B(\boldsymbol{ r})$ is non-constant, or one may have $T_{\perp,0} \neq T_{\|,0}$ if $B(\boldsymbol{r})=B_0$ is a constant. Let us start by investigating the former case. 
\subsubsection*{Spatially varying magnetic field}
For the following calculation we have set $T_{\|,0}=T_{\perp,0}$ everywhere. The varying part of the distribution function, $\delta f$, is given in Eq. \eqref{eq: delta f bimaxwellian}. 
The phase-space coordinates are chosen to be $\boldsymbol{ x} = (\boldsymbol{ z},\mu) = (\boldsymbol{ r},v_\parallel, \mu)$, and the velocity-space volume element becomes
	$$ d\boldsymbol{ v} = \frac{2 \pi B}{m} dv_\| d\mu. $$
The density of states is now equal to
	$$ \Omega'(w,\mu) = \int \delta \left( w - \mu B - \frac{mv_\|^2}{2} \right) 
	\frac{2 \pi B}{m} \; d\boldsymbol{ r} dv_\| = \int_{w>\mu B(\boldsymbol{ r})} \frac{4 \pi B \; d\boldsymbol{ r}}{\sqrt{2 m^3 (w - \mu B)}}, $$
and Eq. (\ref{approximate delta F}) becomes 
\begin{equation}
    \delta F (w,\mu) = \frac{f_0(w)}{\Omega'(w,\mu)} \int \left[ \nu + \tau_\| \left( \frac{w - \mu B}{T_{0} } - \frac{1}{2} \right) + \tau_\perp \left( \frac{\mu B}{ T_{0}} - 1 \right) \right] \frac{4 \pi B \; d\boldsymbol{ r}}{\sqrt{2 m^3 (w - \mu B)}}.
    \label{eq: delta F with mu and nonconstant B}
\end{equation}
If the magnetic field varies with $\boldsymbol{ r}$ in a way that is correlated with the density or temperature fluctuations, then $\delta F$ is generally non-zero and the available energy will then be less than the Helmholtz energy, $A \le H$.

\subsubsection*{Constant magnetic field}
If the magnetic field is constant, $B=B_0$, the perpendicular and parallel temperatures $T_{\perp,0}$ and $T_{\|,0}$ can be different, and will now be taken to be unequal. Under such circumstances, the fluctuating part of the bi-Maxwellian distribution (\ref{eq: bimaxwellian}) with $\boldsymbol{ u}=0$ becomes
\begin{equation}
    \delta f = \left[ \nu + \tau_\| \left( \frac{mv_\|^2}{2T_{\|,0} } - \frac{1}{2} \right) + \tau_\perp \left( \frac{\mu B_0}{ T_{\perp,0}} - 1 \right) \right] f_0,
\end{equation}
where we have perturbed the density and temperatures in the usual manner. The ground state simplifies to
$$ 
\delta F (w,\mu) = f_0 \left[  \langle \tau_\|  \rangle \left( \frac{w - \mu B}{T_{\|,0} } - \frac{1}{2} \right) + \langle \tau_\perp \rangle \left( \frac{\mu B}{ T_{\perp,0}} - 1 \right) \right] ,
$$
and we have employed $\langle \nu \rangle = 0$. The available energy can now be calculated, and becomes
\begin{equation}
    A = \frac{n_0 T_{\|,0} V}{2} \left\langle \nu^2 + \frac{1}{2} \left( \tau_\|^2  - \langle \tau_\| \rangle^2 \right) + \tau_\perp^2 - \langle \tau_\perp \rangle^2 \right\rangle.
    \label{eq: available energy bimaxwellian with constant magnetic field}
\end{equation}
In order to compare this expression to the case without invariants given in Eq. \eqref{eq: AE bimaxwellian no invariants}, we set $T_{\perp,0}=T_{\|,0}=T_0$ in Eq. \eqref{eq: available energy bimaxwellian with constant magnetic field}. Under these conditions, the available energy becomes
\begin{equation}
    A = H - \frac{n_0T_0V}{2}\left( \frac{1}{2} \left \langle \tau_\| \right \rangle^2 + \left \langle \tau_\perp \right \rangle^2 \right).
\end{equation}
We thus conclude that the available energy of an isotropic plasma with constant magnetic field, where the perturbations furthermore satisfy $\tau_{\perp} = \tau_{\|}=\tau$, is equal to the usual Helmholtz free energy if $\mu$ is conserved, and the introduction of anisotropic perturbations reduces it below this value unless $\langle \tau_\| \rangle = \langle \tau_\perp \rangle = 0$. 

\subsection{Magnetic-moment conservation of a kappa-Maxwellian with fluctuating density and temperatures}
We next consider a plasma distribution function that exhibits non-Maxwellian behaviour in the direction along a constant magnetic field $B=B_0$, but is Maxwellian in the perpendicular directions. In order to model such effects, we employ the kappa-Maxwellian distribution function introduced by \cite{hellberg2002generalized},
\begin{equation}
    f_\kappa(\boldsymbol{r},\boldsymbol{v}) = n \left( \frac{m}{2 \pi \overline{T}} \right)^{3/2} \frac{\Gamma(\kappa + 1)}{\kappa^{3/2}\Gamma\left(\kappa - \frac{1}{2}\right)} \sqrt{\frac{\kappa}{\kappa - \frac{3}{2}}} \left( 1 + \frac{m v_\parallel^2}{2 T_\parallel (\kappa - \frac{3}{2})} \right)^{-\kappa} e^{ - \mu B_0 / T_\perp},
\end{equation}
which shares many features with the distribution function discussed in Sec. \ref{sec: kappa distribution in any dimension}. In the limit $\kappa \rightarrow \infty$ it becomes bi-Maxwellian, and at high parallel velocities it exhibits power-law behaviour, $f_\kappa \propto v_\|^{-2\kappa}$. The normalisation is chosen such that the particle number density is equal to $ n $, and the perpendicular and parallel energy densities are 
    $$\int \mu B_0 f_\kappa d \boldsymbol{v} = n T_\perp,$$ 
    $$\int \frac{m v_\|^2}{2} f_\kappa d \boldsymbol{v} = \frac{n T_\|}{2}. $$
One may verify that the spatially constant distribution function is a ground state by setting $m v_\|^2/2 = \epsilon - \mu B_0$ and taking the derivative 
\begin{equation}
    \left( \frac{\partial f_\kappa}{\partial \epsilon} \right)_{\mu,\boldsymbol{r}} =-\frac{f_\kappa \kappa}{\epsilon - \mu B_0 + T_\| \left( \kappa - \frac{3}{2} \right)}.
    \label{eq: monotonicity of kappa-dist}
\end{equation}
Hence it is seen that $f_\kappa$ is a decreasing function of particle energy at fixed $\boldsymbol{r}$ for all $\mu$ as long as $\kappa>3/2$. It follows that any spatially constant $f_\kappa$ is then indeed a ground state. We now slightly perturb the distribution function, so that the fluctuating part becomes
$$
    \delta f_\kappa = f_{\kappa,0} \left[ \nu + \tau_\perp \left( \frac{\mu B_0}{T_{\perp,0}} - 1 \right) + \tau_\| \left( \kappa - \frac{1}{2} - \frac{T_{\|,0} \kappa (\kappa - \frac{3}{2})}{\epsilon - \mu B_0 + T_{\|,0}\left( \kappa - \frac{3}{2} \right)} \right) \right].
$$
The ground state may readily be calculated, and becomes
$$
\delta F_\kappa = f_{\kappa,0} \left[ \left\langle \tau_\perp \right \rangle \left( \frac{\mu B_0}{T_{\perp,0}} - 1 \right) + \left\langle  \tau_\| \right \rangle \left( \kappa - \frac{1}{2} - \frac{T_{\|,0} \kappa (\kappa - \frac{3}{2})}{\epsilon - \mu B_0 + T_{\|,0}\left( \kappa - \frac{3}{2} \right)} \right) \right].
$$
The integrals required for the available energy can be performed analytically, and result in
\begin{equation}
    A = \frac{n_0 T_{\|,0} V}{2 \kappa} \Bigg\langle \frac{1}{2} (\nu + \tilde \tau_\|)^2 + \left( \kappa - \frac{3}{2} \right) \left( \nu^2 + \frac{\tilde \tau_\|^2}{2} \right) 
    + (\kappa - 1) \tilde \tau_\perp^2 \Bigg\rangle,
\end{equation}
where $\tilde \tau_\| = \tau_\| - \langle \tau_\| \rangle$ and $\tilde \tau_\perp = \tau_\perp - \langle \tau_\perp \rangle$. As required, this result is positive definite and becomes equal to Eq. \eqref{eq: available energy bimaxwellian with constant magnetic field}  in the Maxwellian limit, $\kappa \rightarrow +\infty$.

\subsection{Conservation of magnetic moment and the parallel invariant}

For instabilities and turbulence with wavelengths comparable to the ion gyroradius in the direction perpendicular to the magnetic field, the frequency is usually much smaller than that of the motion of electrons along field. The electron distribution function is then independent of the position along the field line in each trapping well. Moreover, since the bounce frequency of magnetically trapped electrons is much larger than that of the fluctuations, the action integral taken between two consecutive bounce points,
    $$ J =  \int_{l_1}^{l_2} m v_\| \; dl, $$
is conserved in addition to the magnetic moment. The available energy of the electrons in a thin flux tube aligned with the magnetic field under the constraint of constant $\mu$ and $J$ was recently calculated by \citet{helander_2020} and \citet{mackenbach_2022,Mackenbach_Proll_Wakelkamp_Helander_2023}. It is interesting to note that their results do not agree with the formulas ({\ref{A1}}) or (\ref{A2}). 

The reason can be traced back to an implicit assumption made in the previous section, where it was tacitly assumed not only that $\delta f \ll f_0$ but also that all derivatives of $\delta f$ are smaller than the corresponding ones of $f_0$. This assumption is, in general, violated if the number of conserved quantities $\boldsymbol{ y}$ is so large that only a single coordinate in phase space is not conserved, i.e., if the vector $\boldsymbol{ z}$ only consists of a single component $z$\footnote{In the calculation by \cite{helander_2020}, the distribution function depends on the phase-space coordinates $(\psi,\mu,J)$, where $\psi$ labels different flux surfaces, and $\mu$ and $J$ are conserved. In this case, we thus have $\boldsymbol{y} = (\mu,J)$ and $z=\psi$. A similar situation arises in the related calculation by \citet{mackenbach_2022} of the available energy in a flux tube, where the latter is taken to be so slender that the dependence on the coordinates $(\psi,\alpha)$ can be taken to be linear in these variables.}. If this is the case, that component can be expressed as a function of the energy and $\boldsymbol{ y}$, at least locally, by inverting the function $\epsilon(\boldsymbol{ y},z)$, and it must therefore be possible to write $\delta f$ as a function of $\boldsymbol{ y}$ and $\epsilon$,
    $$ \delta f (\boldsymbol{ y},z) = \delta F[\epsilon(\boldsymbol{ y},z),\boldsymbol{ y}], $$
where it is expected that $\delta F \ll F_0$. However, since the function
    $$f(\boldsymbol{ y},z) =F_0[\epsilon(\boldsymbol{ y},z),\boldsymbol{ y}] + \delta F[\epsilon(\boldsymbol{ y},z),\boldsymbol{ y}] $$
only depends on $z$ through $\epsilon(\boldsymbol{ y},z)$, it follows that it actually represents a ground state if
    $$ \frac{\partial ( F_0 + \delta F)}{\partial \epsilon} < 0. $$
The perturbed state $f$ is thus a a ground state unless $|\partial \delta F/ \partial \epsilon| > |\partial F_0 / \partial \epsilon|$, which invalidates the perturbation theory developed in Sections 2 and 3. More generally, in order to verify the correctness of the perturbation theory, one could check that that $\partial \delta f / \partial x_i \ll \partial f_0 / \partial x_i$ as required. 
\par 
As a final observation, note that, in the perturbative framework described in Section 3, the available energy is invariant under the substitution $\delta f \rightarrow -\delta f$, as may be verified from Eqs. \eqref{approximate delta F} and \eqref{A1}. For a fixed background magnetic field, then, the available energy (\ref{A1}) is impervious to the relative sign of the gradient of the magnetic field strength, $\nabla B$, and the gradient of $\delta f$. For instance, if $f=f_0 + \delta f$ is a Maxwellian with a small density gradient contained in $\delta f$, then Eq.~(\ref{A1}) is independent of the sign of $\nabla n \cdot \nabla B$. There is thus no notion of ``good'' or ``bad'' curvature, and the expressions (\ref{A1}) and (\ref{A2}) are oblivious to curvature-driven modes although available energy can, in fact, be used to describe trapped-electron modes if $\mu$ and $J$ are treated as adiabatic invariants \citep{helander_2020,mackenbach_2022,Mackenbach_Proll_Wakelkamp_Helander_2023}.

\section{Conclusions}

The present paper can be seen as a continuation of the discussion by \cite{helander_2017,helander_2020} of the available energy in a magnetically confined plasma. 
Here, we have derived explicit formulas, given in Eqs.~(\ref{A1}) and (\ref{A2}), for the available energy of a distribution function $f$ close to a `ground state' $f_0$, a state whose energy cannot be lowered by Gardner restacking that keeps the invariants $\boldsymbol{y}$ constant. In the special case of  a Maxwellian plasma with small density and temperature fluctuations and no adiabatic invariants, the available energy (\ref{eq: AE bimaxwellian no invariants}) was calculated already by \citet{helander_2017}, but his calculation, which proceeded directly from the integral equation for the ground state, is rather complicated. Here, we have found a simpler and more general way, which also allows for the conservation of adiabatic invariants. The result shows explicitly how the available energy is related to Helmholtz free energy, which has recently been used to derive upper bounds on the linear and nonlinear growth of gyrokinetic instabilities \citep{helander_plunk_2022,plunk_helander_2022,Plunk_Helander_2023}. The available energy is given by expression (\ref{A1}) and is closely related to the free energy, which it can never exceed according to Eq.~(\ref{A2}). The Helmholtz free energy is thus an upper bound on the available energy, and it becomes equal to the latter (to leading order) if the ground state $f_0$ is accessible from $f$ through Gardner restacking. 
\par 
These results have been used to explicitly calculate the available energy in a number of special cases: The available energy of a Maxwellian with anistropic temperature fluctuations and no further constraints was shown to be equal to the Helmholtz free energy, whereas invoking invariance of $\mu$ decreases it below this value. The available energy of various non-Maxwellian distribution functions with power-law tails was considered too, and display a dependence on the power-law considered. Finally, in calculations of the available energy in plasmas with a sufficiently large number of conserved quantities, it was shown that the considered asymptotic framework is invalid and the available energy instead needs to be calculated as by \cite{helander_2020} and \cite{mackenbach_2022}. 

Our discussion has been limited to the simplest case of a single particle species, but it would be valuable to extend it to several species which together satisfy the requirement of quasineutrality. 

\section*{Acknowledgements}
We are indebted to three anonymous referees and to Prof. Alexander Schekochihin for several insightful remarks, which have helped to improve the paper. This work has been carried out within the framework of the EUROfusion Consortium, partially funded by the European Union via the Euratom Research and Training Programme (Grant Agreement No 101052200 — EUROfusion), and partly supported by a grant from the Simons Foundation (560651, PH). The Swiss contribution to this work was funded by the Swiss State Secretariat for Education, Research and Innovation (SERI). Views and opinions expressed are however those of the author(s) only and do not necessarily reflect those of the European Union, the European Commission or SERI. Neither the European Union nor the European Commission nor SERI can be held responsible for them. This publication is part of the project ``Shaping turbulence – building a framework for turbulence optimisation of fusion reactors'', with Project No. \texttt{OCENW.KLEIN.013} of the research program ``NWO Open Competition Domain Science'' which is financed by the Dutch Research Council (NWO).

\section*{Declaration of interests}
The authors report no conflict of interest.

\appendix 

\bibliographystyle{jpp}

\bibliography{AE3}

\end{document}